# Secure Links: Secure-by-Design Communications in IEC 61499 Industrial Control Applications

Awais Tanveer✉, Roopak Sinha & Matthew M. Y. Kuo

*Software Engineering Research Laboratory*
*School of Engineering, Computer and Mathematical Sciences*
*Auckland University of Technology (AUT)*
*Auckland 1010, New Zealand*
awais.tanveer@aut.ac.nz; roopak.sinha@aut.ac.nz; matthew.kuo@aut.ac.nz

**Abstract**

*Increasing automation and external connectivity in industrial control systems (ICS) demand a greater emphasis on software-level communication security. In this article, we propose a secure-by-design development method for building ICS applications, where requirements from security standards like ISA/IEC 62443 are fulfilled by design-time abstractions called secure links. Proposed as an extension to the IEC 61499 development standard, secure links incorporate both light-weight and traditional security mechanisms into applications with negligible effort. Applications containing secure links can be automatically compiled into fully IEC 61499-compliant software. Experimental results show secure links significantly reduce design and code complexity and improve application maintainability and requirements traceability.*

## 1. INTRODUCTION

Application-level communication security has become a key concern in industrial control systems (ICS) that control critical systems like smart grids, manufacturing plants, and nuclear facilities. ICS run highly distributed software applications deployed on resource-constrained devices. Combined with increasing automation and external connectivity, these factors make ICS very vulnerable to security attacks. Malicious disruption of ICS can be extremely expensive [1], which necessitates the use of *secure-by-design* [2] approaches for developing ICS applications. Modern ICS must adhere to security standards such as ISA/IEC 62443 [3], common criteria, and NIST SP 800-82. ICS in critical infrastructure are increasingly being constructed through the reuse of certified secure components [4]. These trends mean that current methodologies for developing ICS applications must urgently move from a "security as an after-thought" approach to a systematic, secure-by-design mindset [5]. ICS applications are deployed onto multiple devices and involve extensive communication between devices. Frequent reconfiguration is supported by standards like IEC 61499 [6], but current approaches lack a secure-by-design approach to communication security.

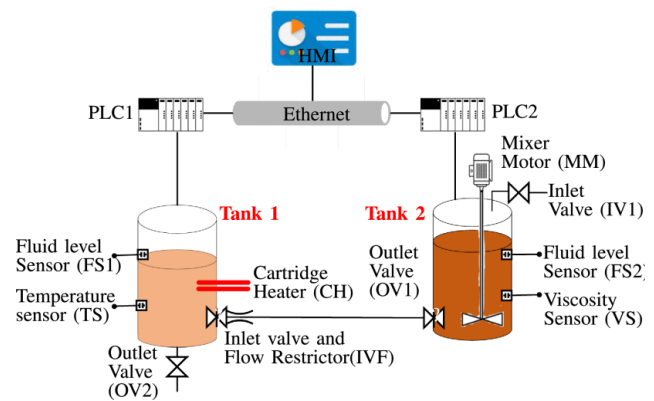

Fig. 1. Industrial mixer control system.

We use the industrial mixer control system (IMCS), shown in Fig. 1, to highlight the communication security challenge in ICS and subsequent concepts introduced in this article. The IMCS is a distributed IEC 61499 application executing on two programmable logic controllers (PLCs) PLC1 and PLC2. The PLCs control the sensors and actuators associated with tanks 1 and 2, respectively. Tank 2 is responsible for maintaining a desired viscosity and level (volume) of a fluid solution. The (part of the) application running on PLC2 controls the inlet valve IV1 through which the solution enters the tank, the mixer motor MM to carry out the mixing, and the outlet valve OV1. The application uses readings from the level sensor FS2 and the viscosity sensor VS to ensure correct operation. Tank 1 heats the fluid solution to the desired temperature. The application running on PLC1 controls the inlet valve and flow restrictor IVF to allow the premixed solution from tank 2 to enter tank 1, the cartridge heater CH to heat the solution, and the outlet valve OV2. The application uses readings from the level sensor FS1 and the temperature sensor to ensure correct operation.

Safe operation of the IMCS requires coordination between the applications running on the two PLCs. When PLC1 detects that more fluid is required in tank 1, it opens the IVF valve and sends a fluid acquisition request to PLC2 for it to release the OV1 valve. When the fluid in tank 2 is ready, PLC2 opens the OV1 valve to move the fluid from tank 2 to tank 1. It is vital that OV1 is only released after IVF to ensure that no fluid is left sitting in the pipe between tank 2 and tank 1. If the fluid is not mixed or heated for a prolonged time, it can harden and cause blockages. The PLCs also interact with an HMI to provide a real-time view of the system.

The IMCS contains typical security vulnerabilities found in ICS. In a *replay attack*, an attacker is able to replay messages between the PLCs. It can request PLC2 to open OV1 when IVF is still closed, causing the fluid solution in the pipe to harden and cause blockages. The IMCS is also vulnerable to a *man-in-the-middle attack* [7], where an attacker intercepts the messages between the PLCs and/or the HMI. It can starve tank 1 by blocking all OV1 release requests from PLC1 to PLC2. Alternatively, the attacker can send sporadic false commands to the PLCs to cause further disruptions or unsafe operation. Such attacks affect the safety and availability of the IMCS, potentially causing physical damage to the system itself and disrupting other systems dependent on the IMCS.

Traditional, widely used approaches for communications security in other domains are not well-suited for the use in ICS due to the security-performance tradeoff [8]. Approaches like transport layer security (TLS), Datagram TLS, and virtual private network (VPN) induce higher communication latency, which may prevent ICS applications from meeting real-time requirements. Second, methods like TLS, for example, are point-to-point secure communication protocols more suitable for client-server architectures. ICS, on the other hand, tend to use unidirectional and multicast communications between application slices. Third, traditional methods may require more resources than are available on resource-constrained PLCs. Subsequently, maintaining a balance between the demands of communication security in ICS becomes a key challenge [9]. Hence, for an ICS application developer, the only available option to protect against such attacks is to include communication security mechanisms manually into the application. This approach does not allow applications to be easily reconfigured, sacrificing a key feature of ICS. For the IMCS, while one can use TLS to secure all communications between the two PLCs, the overhead introduced by this approach may affect the safety of the system, which is dependent on real-time responsiveness. For example, depending on the use of a primitive traditional method may result in missing the threshold time for opening/closing of IVF and OV1 valves that may cause significant damage to the IMCS by overfilling tank 1.

Developing standards-compliant secure ICS applications has further challenges. Standards like ISA/IEC 62443 provide security requirements that must be met by certified systems. Certification requires a rigorous and expensive process of ensuring the system has met each security requirement. The precise set of requirements to be met depends on factors like target security levels and the security-performance tradeoffs. Hence, a one size fits all approach, such as TLS, is rarely a feasible solution. Unfortunately, there is currently no systematic way to *trace* or link requirements from standards to parts of code that they are addressed in [10], making it extremely difficult to certify ICS applications that are often reconfigured and have real-time performance requirements.

We propose a secure-by-design approach for developing ICS applications called the *Secure Links Development Method* (SLDM). SLDM, shown in Fig. 2, as a UML component diagram, is a systematic process to develop secure ICS applications compliant to standards like IEC 62443. SLDM has been designed to adhere to two major security development lifecycles: Microsoft's security development method [11] and the secure product development lifecycle requirements process from IEC 62443-4-1 [12]. To manage requirements from security standards, SLDM uses the Traceability Of Requirements Using Splices (TORUS) framework [13], presented in Section 3-A, which provides a systematic way to organize requirements from a system's cybersecurity requirements specification (CSRS) [14] and links them to application designs and code that implement them. Next, we introduce design abstractions called *secure links*, presented in Section 3-B, which help include communication security mechanisms into IEC 61499 applications in a uniform, consistent, maintainable, traceable, and reusable manner. Secure links contain references to requirements from the TORUS repository and reusable communication security mechanisms that can help meet these requirements. SLDM enables designers to flexibly select the most suited secure communication mechanisms for each part of their ICS applications, from lightweight cryptography to traditional TLS-based approaches, depending on target security requirements. It prevents the need to manually code security mechanisms, resulting in significant reductions in developing secure applications. SLDM is fully compatible with the traditional way of constructing ICS applications so that security mechanisms can also be manually coded into an application, in addition to using secure links when developing secure-by-design applications.

A secure link compiler, presented in Section 4, converts application designs into a fully IEC 61499-compliant code that can be deployed onto target devices like PLCs. The compiler transforms secure links into code for the chosen security mechanisms by using preverified implementations of these mechanisms from a security library. The compiler considers the current configuration of the ICS application to secure only those communications that happen between parts of the application that are deployed over different devices. When the configuration changes, the compiler can regenerate an efficient code that is suitable for meeting the updated communication security needs. The code generated by the compiler retains its link to the requirements in the CSRS, considerably reducing the effort required for certification. Experimental results, presented in Section 5, reinforce that updating, adding, and deleting secure links in an application requires a negligible effort, and applications remain highly maintainable. On the other hand, when the same systems are developed using a traditional approach, each additional security mechanism introduced into the software causes a significant increase in software complexity and a similar decline in maintainability.

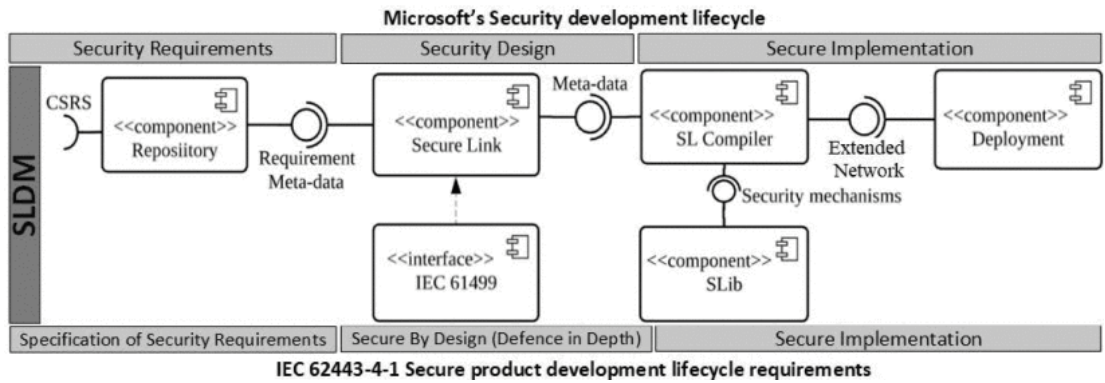

Fig. 2. Overview of the SLDM.

## 2. RELATED WORKS

Recently updated security standards such as ISA/IEC 62443 have become important cornerstones of ICS security in the industry. The recently formed global cybersecurity alliance (GCA) comprising of leading industrial players like Schneider, Rock-well, and Honeywell Automation has decided to use ISA/IEC 62433 as an interchangeable medium to share security-related knowledge and expertise with end-users, manufacturers, and government agencies [15]. Certification schemes for ISA/IEC 62433, such as embedded device security assurance and component security assurance, have also gained popularity, resulting in several devices from large-scale industrial manufacturers being certified [16]. Moreover, the manufacturing industry is rapidly moving toward Industry 4.0 [17] that extensively utilizes ICS. Standardization efforts of cybersecurity requirements for such an environment have focused on the introduction of new standards as well as the adaption of more generic standards like ISO 27001. Some limitations of industry 4.0 and IoT compatible security standards like ISO 27001, IEC 62443, ETSI TS 103 645, and the recommendations by the European Union Agency for Network and Information Security, have been discussed in [18]. The study finds that the implementation of these standards is complex and time/resource intensive, and requires better processes and frameworks in the future.

Several existing works support organizing requirements in a systematic manner. Microsoft's security development lifecycle [11] provides a generic approach to integrate security requirements into software development. IEC 62443-4-1 (secure product development lifecycle requirements) [12] provides a more ICS-specific approach. Other works like the security requirements engineering process [19] use standards like common criteria for eliciting requirements. In [20], relational databases are used to organize requirements from standards for the use during development. In [19], a security resources repository is integrated with requirements from common criteria. In [21], a repository is used for requirements reuse. These existing approaches either rely heavily on textual requirements or do not specify a storage method for requirements. In [22], a survey of different mechanisms to create and maintain traceability in PLC and SysML models is presented, which highlights the need for automatically creating and maintaining traceability between requirements and system components. TORUS [13] presents an ICS-specific approach toward requirements traceability, but it has not been used for security.

Several existing frameworks offer a structured approach to building ICS applications. A design framework that inherently supports formal validation and verification of automation applications, through integrating computer-aided design, unified modeling language, MATLAB/Simulink, and IEC 61499 is proposed in [23]. Another design and development framework, called "methodology for industrial automation systems" [24], generates analysis and design documentation of functional and nonfunctional requirements as structured documents and use-case diagrams, subsequently used to generate code structures. This framework, however, lacks traceability support. vueOne [25] is an end-to-end IDE for developing Industry 4.0 applications, but it does not support requirements or their traceability. In [26], a model-based engineering workflow for IEC 61131-based distributed manufacturing automation systems is presented. Requirements modeled using SysML are traced through the design and development phases. However, the overall approach does not support security engineering.

Secure communication is of particular concern in IEC 61499 applications. Publisher/subscriber models for unidirectional communications use the multicast mode of communication according to conformance profiles like HOLOBLOC and protocols like PROFINET. Bidirectional communications using the client/server model may use TLS between application slices deployed onto multiple devices. However, secure multicast communication is challenging in ICS when using Datagram TLS or IPSEC-based virtual private networks [9]. Other challenges of applying TLS may include performance tradeoffs, channel multiplexing, constrained components, time synchronization. There is a strong case of using OPC-UA security for ICS communication; however, its implementation in an ICS is not straightforward, mainly due to the large resource footprint and lack of real-time capabilities [27]. In an ICS, not all communications need securing, with partial protection of data being sufficient for many applications [28]. Selective protection of "data of interest" helps in better system performance while ensuring

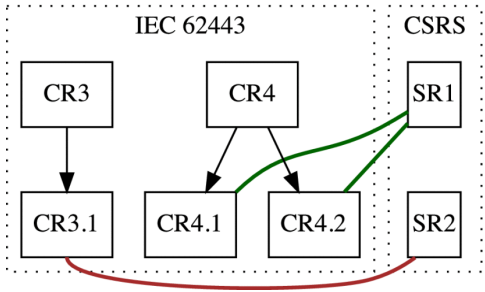

Fig. 3. TORUS-based CSRS repository.

adequate security. In such cases, a blanket approach like TLS may induce latency overheads [9].

*Lightweight cryptography* includes transforms for resource-constrained devices featuring better memory footprints and performance than traditional cryptographic primitives. Ongoing efforts to standardize lightweight security mechanisms include NIST's new lightweight cryptography standardization selection process [29]. Mainstream IT system security solutions like TLS and IPSEC do not contain lightweight ciphers [8]. Lightweight VPNs such as wireguard [30] are expected to find use in industrial settings in the future.

Current security standards provide comprehensive sets of requirements. However, in the absence of processes that integrates a security focus and the development process, the implementation of such standards remains challenging. Standalone methods and framework discussed above are disjointed and do not provide a comprehensive end-to-end engineering solution for secure-by-design ICS applications. Current literature reports gaps like a lack of security standard requirements specification, end-to-end requirements traceability, and change management. Development standards like IEC 61499 do not inherently support secure-by-design constructs. Moreover, the IEC 61499 network communication models and constrained ICS environments pose a significant challenge in balancing security and performance. This article proposes a process for developing secure-by-design ICS applications, which alleviates several gaps in current research.

## 3. SECURE LINKS FOR APPLICATION DESIGN

### A. Security Requirements Repository

IEC 62443-4-2 "Security for industrial automation and control systems—Part 4-2: Technical security requirements for ICS components" [31] provides generic requirements for achieving specific *security levels*. Security requirements from possibly multiple standards are organized in a CSRS. Consider, for example, two requirements for the IMCS case study.

1) *SR*1:*preventdisclosureofcriticalparameterstoanunauthorized party during transmission on external interfaces.* SR1 is linked to *Foundation Requirement 4* of IEC 62443 relating to data confidentiality, and more specifically to IEC 62443-4-2 component requirements (CRs) CR4.1 (Information confidentiality) and CR4.3 (Use of cryptography), for security level L4.

2) *SR2: ensure appropriate data integrity of critical parameters during transmission on external interfaces.* SR2 is related to *Foundation Requirement 3, CR3.1* of IEC 62443-4-2 that deals with communication integrity.

We use the TORUS [13] framework to organize security requirements and achieve traceability (illustrated as phase 1 in Fig. 2). Fig. 3 shows how requirements are organized in a TORUS-based CSRS repository. The cluster labeled as "IEC 62443" shows how TORUS uses a graph structure to retain the hierarchy of requirements from a security standard. The cluster labeled "CSRS" shows the references to system requirements. Cross-linkages or *splices* are used to connect generic requirements from the standard, such as *CR4.1* and *CR4.2*, to system-specific requirements, such as SR1. Requirements engineers first create these splices. A TORUS-based requirement repository RR can be seen as a set of requirements $\{r_1, r_2, \ldots r_n\}$. For Fig. 3, RR = {SR1, SR2}.

TORUS provides a number of useful features for traceability. Each requirement in the repository can be annotated with additional meta-data including its *scope* (system or component) and the target security-level. SLDM is concerned only with communications security and, therefore, operates on a subset of the requirements in the CSRS. These requirements tend to be a component-level in nature as they relate to the interaction between components. TORUS can automatically update splices when the documents containing standard and/or the system-specific requirements change. TORUS provides graph algorithms to determine important metrics like coverage, which are described in more detail in [13]. As we describe later in Section 4, TORUS is used to trace standards-based requirements all the way to the deployed code, which significantly reduces security standard compliance costs.

### B. Secure Links

Adding security mechanisms to an ICS application requires using the same constructs, such as function blocks (FBs) in IEC 61499, that are used to build the control logic and algorithms. While this provides familiarity, it can obscure the boundaries between security mechanisms and other parts of the application, affecting code maintainability, application reconfigurability, and standard certification costs. In this section, we propose *secure links*, which are reusable design abstractions for building secure-by-design IEC 61499 applications. Secure links allow designers to flexibly include both lightweight and traditional methods like TLS to secure different parts of an ICS application, depending on the target security requirements and considering the security-performance tradeoff. Secure links extend *FB networks*.

**Definition 1. FB Network:** An FB network is defined as a pair $fbn = (FB, C)$, where FB is a set of FB instances. For each $fb \in FB$, we define the following.

1) $fb.IE$ and $fb.OE$ are sets of event inputs and outputs.

2) $fb.IV$ and $fb.OV$ are sets of data inputs and outputs.

3) An association function $fb.\mathcal{A}: fb.IV \cup fb.OV \rightarrow fb.IE \cup fb.OE$ maps each input/output variable to a unique input/output event associated with it.[1]

$C$ is a set of connections. Each connection $c \in C$ is described as a pair $(src, trg)$ such that the following hold.

1) $c.src \in \bigcup_{i \in |FB|}(fb_i.OV \cup fb_i.OE) \cup \{\square\}$ is the source.
2) $c.trg \in \bigcup_{i \in |FB|}(fb_i.IV \cup fb_i.IE) \cup \{\square\}$ is the target.

$\square$ indicates an unconnected input or output. Furthermore, $C$ is restricted as follows.

1) For any connection $c \in C$, if $c.src$ is an event (var), then $c.trg$ must also be an event (var) or $\square$, and vice versa.
2) $C$ cannot contain a connection $(\square, \square)$.
3) For any *data* connections $c_1, c_2 \in C$, $c_2.trg \neq c_1.trg$.

Fig. 4 shows the FB network for the IMCS application containing four interconnected FB instances. The instance MTank of type MainTank contains input/output events and variables like INIT, ViscS, INITO, and InReq. FBs contain associations, such as between STank.InReq and STank.IR (not depicted in network diagrams). Unidirectional event and variable connections include (MTank.IR,STank.IR) and (MTank.InReq,STank.InReq). Open connections include (MTankPub.MData,$\square$). Note that top-level FB networks, such as in Fig. 4, are *applications*, which can be deployed onto multiple devices based on a configuration mapping. In Fig. 4, instances MTank and MTankPub are deployed onto PLC1, and remaining blocks are deployed onto PLC2, as illustrated by the coloring of the instances. Formally, given an application $fbn$ and a set D of available resources, the deployment configuration or *mapping* is defined as $\mathcal{D}: fbn.FB \rightarrow D$.

When using secure links, designers choose appropriate security mechanisms for which preverified implementations are available in a *security library*.

**Definition 2. Security Library:** A Security Library $SLib$ is a finite set of *security mechanisms* (FB networks) $\{sm_1, ..., sm_n\}$, such that each $sm \in SLib$ has a signature $sm.S = (cin, cout, params)$ where the following hold.

1) $cin$ is a variable connection where $cin.src = \square$.
2) $cout$ is a variable connection where $cout.trg = \square$.
3) $params \subset \bigcup_{i \in |FB|} fb_i.IV$ is a set of unconnected input variables or parameters.
4) $sm.FB$ is partitioned into sets $sm.FB_a$ and $sm.FB_b$. $sm.FB_a = \emptyset$ implies a

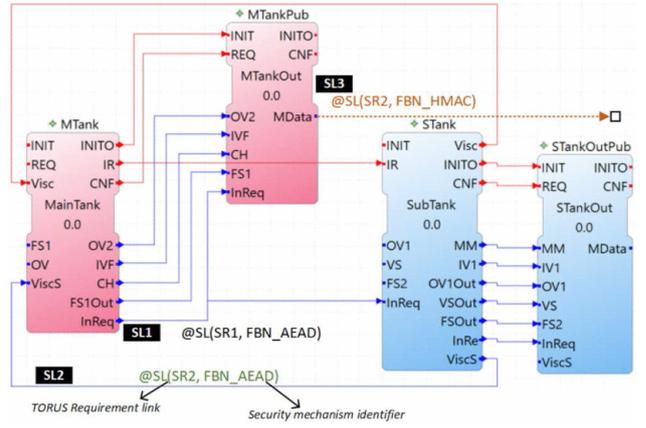

Fig. 4. IEC 61499 implementation of IMCS including secure links. The blocks are colored w.r.t. to device mapping.

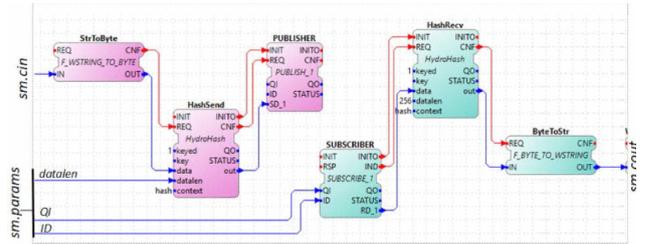

Fig. 5. Send-receive security mechanism for lightweight hashing.

5) receive mechanism where $sm.FB_b = \emptyset$ implies a send mechanism. Otherwise, the mechanism is send-receive.

Fig. 5 shows an example of a security mechanism implementing lightweight secure hashing. The FB instances of the mechanism are partitioned into two sets $FB_a$ and $FB_b$, each containing three blocks, indicated by the coloring of the blocks. As both $FB_a$ and $FB_b$ are nonempty, this is a send-receive mechanism, which requires deployment into two separate devices, responsible for sending and receiving the communications. This partitioning also makes security mechanisms distinctly different from *composite* FBs in IEC 61499, which contain networks that must be deployed on the *same* device. This mechanism contains a set of parameters $params = \{datalen, QI, ID\}$ that can be configured manually or be assigned default values.

Key exchange is an integral part of many security mechanisms, but it becomes increasingly complicated for larger sets of devices. In our case, where needed, each security mechanism includes the required logic for key exchange, in the form of additional FB instances, providing a dedicated (up to) two-device key exchange.

**Definition 3. Secure Links:** For a FB network $fbn = (FB, C)$, $SL \subset C$ is a set of secure data links such that for each $sl \in SL$ contains the following *annotations*.

1) $sl.r \in RR$ is a requirement from a repository $RR$.
2) $sl.sm$ is the reference to a *security mechanism* and $sl.\bar{pv}$ are parameter assignments needed to instantiate $sl.sm$.

---
[1] Variables can be associated with multiple events. Restricting to one-to-one associations is for readability and does not affect the approach's generality.

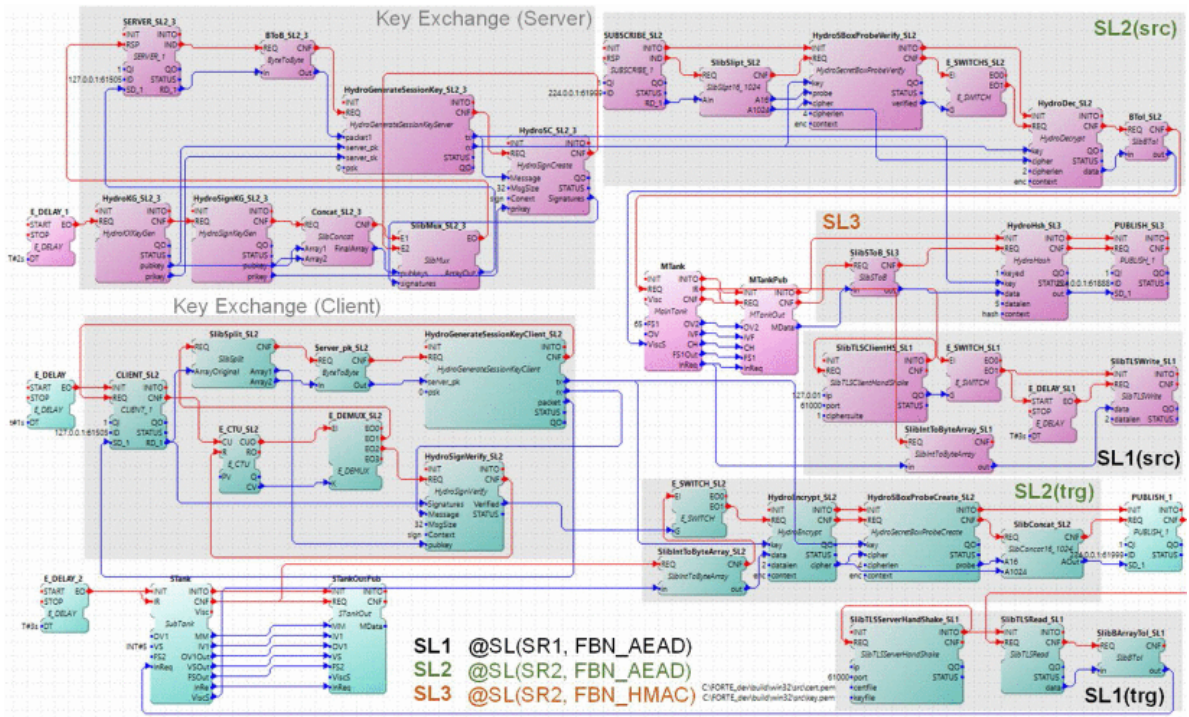

Fig. 6. Fully IEC 61499-compliant IMCS application generated by Algorithm 1.

The use of secure links forms the second part of the SLDM process described in Fig. 2. A possible approach, such as, with TORUS, may involve using a graph database to store the requirement repository $RR$ and the security library $SLib$ that allows designers to interactively query these resources to identify security requirements and then choose a suitable security mechanism to address each requirement. A secure link simply annotates a variable connection of a FB network with two annotations: a requirement from the CSRS stored in the repository $RR$ and a configured security mechanism from the secure library $SLib$. Fig. 4 illustrates the use of secure links in the IMCS application. Consider the data connection/secure link $sl = (MTank.InReq, STank.InReq)$. Its annotations include $sl.r = SR1$, which is one of the security requirements in the TORUS repository, discussed in Section 3-A. Additionally, the second annotation $sl.sm = FBN\_AEAD$ mentions the security mechanism chosen by the designer from the *security library*. The assignment of parameters $sl.\bar{p}v$ allow the designer to configure the mechanism appropriately. Note that configured parameters are not illustrated in Fig. 4 for readability.

SLDM depends on the availability of a comprehensive security library, which does require initial effort to develop. Moreover, the inclusion of newer security mechanisms, such as for active security protection, requires this library to be frequently updated and maintained.

## 4. COMPILING SECURE LINKS INTO IEC 61499-COMPLIANT SOFTWARE

Applications containing secure links can be transformed into IEC 61499-compliant applications that can then be deployed to any PLC. Algorithm 1 shows the steps involved in this transformation. Overall, the algorithm replaces each secure link by a configured instance of the referenced security mechanism (lines 2–22), *only* if the two ends of the link connect FB instances deployed onto different devices (line 3). The chosen mechanism is first retrieved from the security library (lines 4–7) and is added to the application in lieu of the original secure link (lines 7 and 8). The device partitions $FB_a$ and $FB_b$ of the mechanism are mapped to be deployed onto the separate devices hosting the communicating blocks in the input network (lines 10 and 15). The data connection in the secure link is replaced by adding in the mechanism via connections *cin* and *cout* of the network (lines 11 and 16), with a similar adjustment in associated events (lines 12 and 17). The parameters for the mechanism are adjusted according to the values provided by the designer (line 19) or by fetching the stored default values for the parameters. In line 20, the compiler signals to TORUS that the secure link has been replaced by a configure network for the security mechanism, for traceability purposes.

Fig. 6 shows the fully IEC 61499-compliant network obtained via the compilation of the application shown in Fig. 4. The FB network added for every security mechanism is illustrated by a bounding box annotated with the text of the corresponding secure link. $SL1(FB_a)$ and $SL1(FB_b)$ are the send and receive parts of the security mechanism, respectively, providing TLS-based AEAD security. $SL2(FB_a)$ and $SL2(FB_b)$ are the send and receive parts of the security mechanism, respectively, providing lightweight AEAD security. $SL3(FB_a)$ denotes only the sending part of a lightweight hashing security mechanism. A key exchange server and a client are the send and receive parts of a key exchange protocol, which provides key generation for both SL2 and SL3. Generating common key exchanges for secure links distributed over the same pairs of devices is a simple compiler optimization for faster performance. SL1 consumes keys generated during TLS

**Algorithm 1:** Secure Link Compilation.

```
input: Input network fbn_in, Library SLib
output: Transformed network fbn_out
1:  Initialize fbn_out = fbn_in
2:  for all sl ∈ fbn_in.SL do
3:    if (D(fb(sl.src)) ≠ D(fb(sl.trg))) then
4:      Retrieve sl.sm from SLib
5:      Instantiate sl.sm to network fbn_s
6:      Name each instance fb in fbn_s as fb_sl
7:      Add fbn_s to fbn_out
8:      Remove sl from fbn_out
9:      if sl.src ≠ □ then           (receive mechanism)
10:       map fbn_s.FB_a to D(fb(sl.src))
11:       Set fbn_s.S.cin.src to sl.src
12:       Replicate all associations of fbn_in.A(sl.src) in
          fbn_s.A(fbn_s.S.cin.src)
13:     end if
14:     if sl.trg ≠ □ then            (send mechanism)
15:       map fbn_s.FB_b to D(fb(sl.trg))
16:       Set fbn_s.S.cout.trg to sl.trg
17:       Replicate all associations of fbn_in.A(sl.trg) in
          fbn_s.A(fbn_s.S.cout.trg)
18:     end if
19:     assign fbn_s.S.params = sl.p̄v
20:     Signal to TORUS that fbn_s replaces sl
21:   end if
22: end for
return fbn_out
```

handshakes, and so it does not require dedicated key exchange.

The compiler generates additional splices for TORUS to link the generated code directly to security standards. Fig. 7 shows how TORUS uses explicit references to the requirements contained within secure links to automatically create splices between requirements and secure links, shown as linkages between the clusters "CSRS" and "Secure Links" in Fig. 7. During compilation, additional splices between a secure link and the FB instances added to the network are created (line 20 of Algorithm 1). More persistent linkages are created by naming the instances appropriately in line 6 of Algorithm 1. As Fig. 7 shows, during compliance checks, CRs CR4.1 and CR4.2 from IEC 62443 standard can be automatically and directly linked directly to specific FB instances that address them.

Currently, we have developed and tested a Java implementation of the algorithm in an IEC 61499 compliant eclipse-based 4DIAC IDE plugin.

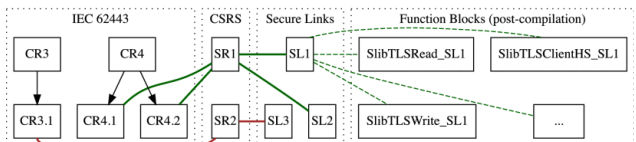

Fig. 7. Tracing requirements from security standards to final implementation using TORUS.

---

[2] [Online]. Available: https://github.com/emsoftaut/securelinks-iec61499

## 5. EXPERIMENTAL RESULTS

Intuitively, using secure links allows for an easier way to construct complex ICS applications with secure-by-design communications. In this section, we quantify the impact of secure links on critical measures like *design complexity* and *maintainability*, which measure the effort required to build and refine secure ICS applications. In his article, we perform this analysis over two case studies: the IMCS case study and a large baggage handling system (BHS) application [32].

Currently, the SLDM design process and compilation are implemented as a single Eclipse plugin for the 4Diac IDE. Designers can add secure link annotations over data connections and then compile designs into fully IEC 61499 and 4DIAC compliant applications. The security library is implemented as a resource library for 4DIAC, and each security mechanism in the library is implemented using standard IEC 61499 features and is stored using the 4DIAC-compatible XML format. SLDM is independent of the IDE used and can be extended to other IDEs in the future. The SLDM plugin also provides an interface to TORUS for end-to-end traceability between requirements from standards and final implementations. Currently, some manual effort is required to invoke TORUS when an artefact (requirement, application design, secure link, or security mechanism) changes. A fully automated interface with TORUS is another important future implementation task for SLDM. All final implementations used in our experiments execute using the FORTE runtime environment and are distributed over Wago PFC200 PLCs. For the IMCS application shown in Fig. 4, the FB instances `MTank` and `MTankPub` execute on PLC1, and `STank` and `STankOutPub` execute on PLC2.[2]

Measuring the complexity of IEC 61499 applications requires adapted measures that provide more accurate results than generic measures like program length. Specifically, we calculate the following three measures, adapted for IEC 61499 and presented in [33], to characterize the complexity of a FB $i$.

1) Structural complexity $S(i) = f_{out}^2(i)$ where $f_{out}$ is the number of outgoing connections or afferent coupling.

2) Data complexity $DC(i) = [NI + NO]/[f_{out}^2(i) + 1]$ where NI and NO are the numbers of data inputs and outputs, respectively.

3) System complexity $C(i) = S(i) + DC(i)$.

Table I. COMPLEXITIES FOR APPLICATIONS AND SECURITY MECHANISMS

| Function block Network | $f_{out}$ | NI | NO | $S^*$ | $DC^*$ | $C^*$ |
|---|---|---|---|---|---|---|
| IMCS | 4 | 29 | 23 | 8 | 29.6 | 37.6 |
| IMCS + secure links | 4 | 41 | 23 | 8 | 41.6 | 49.6 |
| BHS | 48 | 133 | 82 | 196 | 37 | 233 |
| BHS + secure links | 48 | 217 | 82 | 196 | 121 | 317 |
| **Design Complexities for security mechanisms** | | | | | | |
| $sm$ (Key Exchange (L)) | 29 | 54 | 64 | 69 | 29.8 | 98.8 |
| $sm$ (Key Exchange (TLS)) | 4 | 11 | 8 | 8 | 3.8 | 11.8 |
| $sm$ (AEAD (L)) | 17 | 46 | 43 | 27 | 30.1 | 57.1 |
| $sm$ (AEAD (TLS)) | 7 | 14 | 17 | 9 | 14.3 | 23.3 |
| $sm$ (Hash/HMAC) | 6 | 27 | 23 | 6 | 25 | 31 |

Table I describes the complexity values for the IMCS and BHS applications (without and with secure links) as well as for each of the security mechanisms we have implemented into the security library. Table I contains the aggregated values of $S$, $DC$, and $C$ for each network, represented as $S^*$, $DC^*$, and $C^*$, respectively, and defined as the sums of all $S(i)$, $DC(i)$, and $C(i)$ for each FB instance in the network.

Fig. 8 shows the cumulative increase in precompilation complexity when security mechanisms are successively added using secure links or through manual coding. The bars labeled "IMCS" represent the base complexity of the system. "Secure Links" represents IMCS system complexity when secure links are added for three mechanisms: AEAD(TLS), AEAD(L), and Hash. The groups "+AEAD(TLS)," "+AEAD(L)," and "+Hash" show the complexities when the corresponding mechanisms are added successively to the base design represented by the "IMCS" group.

Using secure links has a negligible effect on application complexity. For IMCS, system complexity rises to 49.6 from 37.6. For BHS, system complexity increases to 317 from 233. On the other hand, when all three mechanisms are added manually, the system complexity S∗ for the BHS and IMCS rises from 37.6 to 259.6.6 and 233 to 1198.6, respectively. Note that the complexity, shown by the "+AEAD(TLS)," "+AEAD(L)," and "+Hash" groups is the same as the *postcompilation* complexity of the fully IEC 61499 compliant code generated by the SLDM compiler from the secure links designs represented by the "Secure links" group.

Fig. 8 shows a rise in complexity when (more) security mechanisms are added manually. Mechanisms coded into an application can be either lightweight or traditional. Traditional methods like TLS need support through established protocols such as openSSL, which exist outside of the application. However, this is a blanket approach and does not provide flexibility as *all* interdevice communications are secured instead of only the ones marked secure by the designer. It sacrifices performance, especially in ICS where applications execute on resource-constrained PLCs. In contrast, lightweight methods provide the flexibility of securing a subset of interdevice communications, which results in better performance. However, as Fig. 8 shows, coding the lightweight methods like AEAD(L) adds more complexity to an application than traditional methods because of the larger number of FBs are required to implement such mechanisms completely (and without external infrastructure). The lightweight Hash algorithm is an exception because it causes a relatively smaller increase in complexity. It is because hash functions are inherently simpler to implement, and only one (sending) part of the mechanism is added to each network in our experiments.

Program-level complexity and maintainability can be measured by IEC 61499 relevant measures proposed in [33]. The Halstead's metric for static source complexity involves computing measures $N, n, \hat{N}, PR, V, D,$ and $E$ corresponding to the number of operators and operands, program vocabulary, estimated length, purity ratio, program volume, program difficulty, and program effort, respectively. Due to space limitations, we include the formula used to compute program difficulty $D$, and refer to reader to [33] for details on how other measures are computed. $D$ is computed as

$$D(fbn) = \frac{\sum_{i=1}^{n} D(fb_i)}{n}; D(fb) = \sum_{i=1}^{n} D(\text{alg}) + D(cf).$$

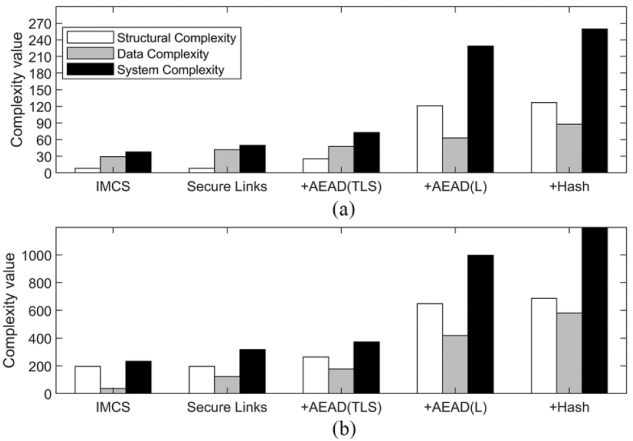

Fig. 8. Cumulative effect on the design complexity of IMCS and BHS after secure links are processed by Algorithm 1.

$D(fbn)$ and $D(fb)$ are the program difficulty of an application and a constituent FB, respectively. Similarly, McCabe's cyclomatic complexity metric $M_M$ is calculated as

$$M_M(fbn) = \frac{\sum_{i=1}^{n} M_M(fb_i)}{n}; M_M(fb)$$
$$= \sum_{i=1}^{n} V(\text{alg}) + V(cf).$$

$M_M(fbn)$ and $M_M(fb)$ represent the cyclomatic complexity of an application and a constituent FB, respectively. $M_M(fb)$ calculations use the value $V$, which is the sum of the cyclomatic complexity of algorithms $V(alg)$ and the cyclomatic algorithms of the control flow $V(cf)$ in a FB. $M_M(fbn)$ is simply the average of all $M_M(fb)$ values in the network.

Table II describes $M_H$ and $M_M$ for the IMCS and BHS applications (without and with secure links) and the security mechanisms we have implemented. To calculate $M_H$ and $M_M$ for applications containing secure links, we consider each link as an operator and its parameters as operands. For example, SL1 secure link annotation is considered as an operator, and the parameters SR1 and *FBN_AEAD* are considered as operands. The MH and MM values are integrated into a single value called the maintainability index (MI).[3] A higher MI value indicates a more maintainable program.

Program difficulty $D$ is a vital metric that has a direct influence on the overall implementation effort of a module. Fig. 9 shows that final cumulative program difficulty and cyclomatic complexity $V$ increase sharply as security mechanisms comprising of individual FBs from Table II are manually added to the IMCS and BHS applications. When three security mechanisms are added, $D$ rises from 11.5 to 178.7 for the IMCS and from 241.7 to 409 for the BHS. Other measures in $M_H$ and $M_M$ show a similar trend of sharp increases for manual coding.

---

[3] Further details of all calculations can be found in [33].

TABLE II. PRECOMPILATION PROGRAM COMPLEXITY AND MAINTAINABILITY

| Name | Halstead's Metric ($M_H$) | | | | | | | McCabe's Metric | | | Maintainability |
|---|---|---|---|---|---|---|---|---|---|---|---|
| | $N$ | $n$ | $\hat{N}$ | $PR$ | $V$ | $D$ | $E$ | $V(alg)$ | $V(cf)$ | $M_M$ | $MI$ |
| IMCS (overall) | 44 | 35 | 110.09 | 5.20 | 183.94 | 11.50 | 1207.80 | 4 | 4 | 8 | 93.51 |
| IMCS+secure links | 62 | 53 | 124.35 | 8.77 | 213.94 | 14.50 | 1222.80 | 7 | 6 | 13 | 89.31 |
| BHS (overall) | 1632 | 272 | 1605.57 | 8.17 | 12101 | 241.79 | 2583135 | 60 | 11 | 71 | 42.19 |
| BHS+secure links | 1650 | 290 | 1619.83 | 11.74 | 12131 | 244.79 | 2583150 | 63 | 14 | 77 | 42.07 |
| Program complexity and maintainability index for cryptographic function blocks of security mechanisms | | | | | | | | | | | |
| HydroKxKeyGen | 51 | 33 | 106.06 | 4.76 | 217.43 | 8.46 | 1351.38 | 7 | 3 | 10 | 82.66 |
| HydroSignKeyGen | 47 | 33 | 107.01 | 4.99 | 199.59 | 6.68 | 930.90 | 6 | 3 | 9 | 85.76 |
| HydroGenerateSessionKeyServer | 70 | 40 | 148.31 | 4.87 | 326.07 | 6.27 | 1490.13 | 8 | 3 | 11 | 77.07 |
| HydroSignCreate | 61 | 40 | 144.99 | 5.15 | 282.35 | 7.41 | 1555.54 | 8 | 3 | 11 | 78.58 |
| HydroGenerateSessionKeyClient | 62 | 41 | 150.86 | 5.22 | 289.75 | 7.13 | 1529.96 | 8 | 3 | 11 | 78.44 |
| HydroSignVerify | 70 | 46 | 178.07 | 5.36 | 342.35 | 8.71 | 2322.81 | 8 | 3 | 11 | 76.10 |
| HydroEncrypt | 97 | 48 | 186.10 | 4.68 | 487.82 | 16.83 | 7080.21 | 11 | 3 | 14 | 71.58 |
| HydroDecrypt | 113 | 52 | 209.57 | 4.63 | 586.42 | 19.20 | 9914.63 | 13 | 3 | 16 | 68.68 |
| HydroSecretBoxProbeCreate | 77 | 43 | 160.08 | 4.84 | 369.59 | 11.09 | 3319.27 | 7 | 3 | 10 | 76.65 |
| HydroSecretBoxProbeVerify | 89 | 45 | 170.85 | 4.67 | 436.50 | 12.79 | 4658.63 | 7 | 3 | 10 | 74.72 |
| HydroHash | 121 | 59 | 251.02 | 4.88 | 654.53 | 21.11 | 12320.04 | 10 | 3 | 13 | 64.35 |
| SlibTLSServerHandShake | 123 | 57 | 246.33 | 4.80 | 658.76 | 12.31 | 6911.19 | 11 | 3 | 14 | 65.91 |
| SlibTLSClientHandShake | 120 | 56 | 238.24 | 4.78 | 638.70 | 13.86 | 7624.16 | 9 | 3 | 12 | 66.53 |
| SlibTLSRead | 54 | 38 | 132.34 | 5.23 | 243.81 | 8.06 | 1457.61 | 4 | 3 | 7 | 83.68 |
| SlibTLSWrite | 55 | 40 | 143.72 | 5.43 | 253.20 | 7.39 | 1373.46 | 4 | 3 | 7 | 84.47 |

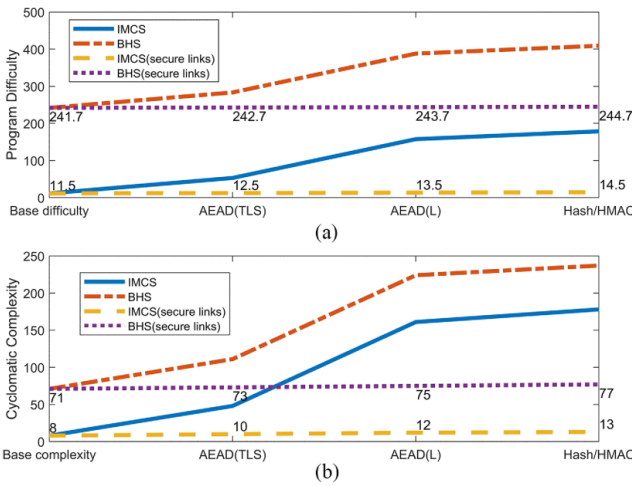

Fig. 9. Manual addition of security mechanisms versus the use of secure links: cumulative program difficulty ($D$) and cyclomatic complexity ($V$) comparisons.

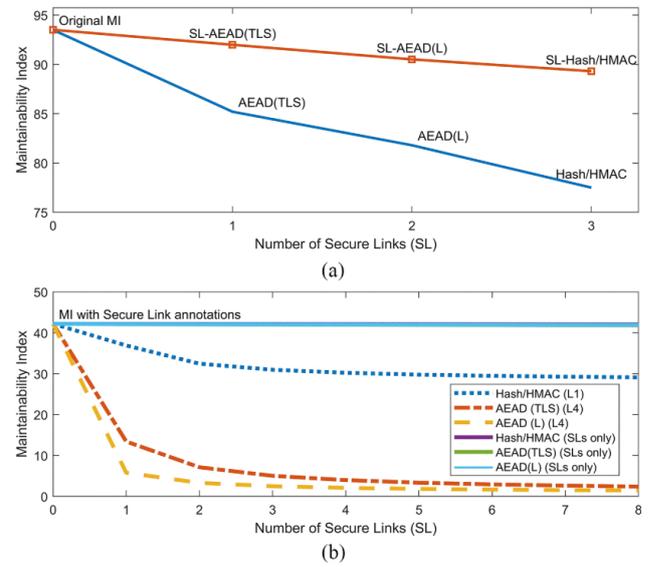

Fig. 10. Maintainability of IMCS and BHS with and without secure links.

Fig. 10 shows a significant decline in maintainability when security mechanisms are added manually. On the other hand, using secure links results in a minimal drop in maintainability. Adding three secure links to the IMCS results in only a slight drop in maintainability from 93.5 to 89.3, which is also described in Table II. The BHS application was tested for multiple security levels provided by IEC 62443-4-2 (CR 3.4). Security level L1 can be achieved by providing integrity, while L2 requires the implementation of integrity and authentication mechanisms such as HMAC. L4 requires encryption for some CRs. Overall, eight different secure links corresponding to varying security levels were added to the BHS. Fig. 10(b) shows that MI decreases from 42.19 to below 20 for L4 when mechanisms are added manually. With secure links, MI sees only a minor drop from 42.19 to 41.9.

As discussed previously, and illustrated in Fig. 8, SLDM supports both lightweight and traditional security mechanisms. Table III describes the average latency of a hundred transactions between the two Wago PFC200 PLCs while using different security mechanisms in the IMCS application. The clocks on both PLCs were synchronized using the network time protocol. We use `libhydrogen` for lightweight mechanisms and `openssl` to support TLS. Both libraries use fast `Curve25519` elliptic curve-based key exchange with `libhydrogen` employing lightweight `Gimli` permutation operations that further suit resource-constrained environments. For all transactions, the lightweight approach takes significantly lesser time than the traditional TLS approach. TLS has a 32% overhead for key exchange over the lightweight method. For Authenticated Encryption with Associated Data (AEAD), `libhydrogen`'s secret box API consumes 3.34 ms as compared to TLS using `AES128-GCM-SHA384`, that has a higher latency of 4.90 ms. Achieving security level L3 in the IMCS requires authentication and integrity with no encryption. In this case, generic hashing with the `libhydrogen`'s key API has an overhead of 1.87 ms, compared to 3.03 ms consumed by TLS (`NULL-SHA256`).

Table III. LATENCY COMPARISON BETWEEN LIGHTWEIGHT AND TLS SECURITY MECHANISMS IMPLEMENTATIONS

|  | Lightweight | TLS |
| --- | --- | --- |
| Key Exchange | 373.27 ms | 492.65 ms |
| AEAD | 3.34 ms | 4.90 ms (AES256, SHA384) <br> 4.80 ms (AES128, SHA256) |
| CBC Mode |  | 4.03 ms (AES128, SHA256) |
| Integrity + Authentication | 1.87 ms | 3.03 ms |

Overall, experimental results show that secure links and the automatic compilation process assist in building applications with low complexity and high maintainability. Although the results are specific to the implementations of TLS and lightweight mechanisms that we have implemented, the use of secure links is independent of the implementations of security mechanisms. SLDM allows adding new mechanisms or optimizing currently included mechanisms at any time without affecting secure-links-based application designs. Our results show that traditional approaches like TLS result in slightly lesser complexity increases than lightweight approaches but have much higher performance overheads and lack the flexibility required in choosing a specific subset of interdevice communications to secure. Manually adding mechanisms, both traditional and/or lightweight, causes a significant drop in application maintainability and a corresponding rise in complexity, which can be avoided by using secure links. Even when a security mechanism is not available in the security library, it is more beneficial to implement it first as a security library element, which can then be used flexibly using secure links. Secure-links-based applications are easier to refactor when security requirements or target security levels change. Additionally, the integration of TORUS into SLDM allows end-to-end traceability between standards and final implementations (as shown in Fig. 7). Finally, secure links are highly reusable design abstractions. The BHS case study implementation contains multiple composite FBs where each block controls a conveyor belt. All such conveyor belt controllers must satisfy the same set of communication security requirements. Hence, a secure link used to secure one such controller can simply be reused for all others.

## 6. CONCLUSION

Secure-by-design approaches can significantly reduce the effort required to build certifiably secure ICS applications. We proposed an ICS development method called the SLDM encapsulating the concept of secure links that are design abstractions supporting the development of secure-by-design IEC 61449 distributed applications. Secure links provided clear traceability between security requirements from ISA/IEC 62443 and application code, which helped reduce the efforts required for certification. Necessary design automation, proposed as a fully IEC 61499-compliant compilation of secure-links-based applications, ensured that very intricate designs could be built with a negligible impact on design-time complexity and application maintainability. Future directions for this work include the development of mature tools that can be integrated into mainstream ICS development workflows and tools, as well as extending the work to include additional ICS security and development standards.

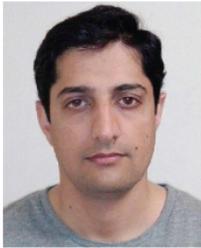

**Awais Tanveer** (Member, IEEE) received the B.E. degree in software engineering from the Bahria University, Islamabad, Pakistan, in 2007, and the M.Sc. (Engg.) degree in software engineering from the University of Engineering and Technology, Taxila, Pakistan, in 2013. He is currently working toward the Ph.D. degree with the Auckland University of Technology (AUT), Auckland, New Zealand.

His current research interests include software requirements engineering and design of cyber–physical systems, industrial automation systems with the particular interest in the aspect of secure communications.

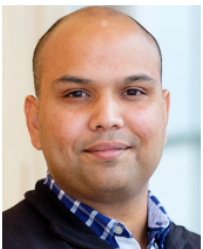

**Roopak Sinha** (Senior Member, IEEE) received the B.E. (Hons.) degree in electronics and computer engineering from the Manukau Institute of Technology, Auckland, New Zealand, in 2003, and the Ph.D. degree in electrical and electronics engineering the M.C.E. degree from The University of Auckland, Auckland, New Zealand, in 2009 and 2016, respectively.

He is an Associate Professor with the School of Engineering, Computer and Mathematical Sciences, The Auckland University of Technology, Auckland, New Zealand. He has previously held research positions with The University of Auckland and INRIA, France. His current research interests include systematic, standards-first design of complex, next-generation embedded software applied to domains like Internet-of-Things, edge computing, cyber–physical systems, home and industrial automation, and intelligent transportation systems. He has served on several international standardisation projects, and works with several New Zealand companies to systematically reduce standards-compliance costs in complex software-driven systems.

Dr. Sinha was named a Senior Fellow of the Higher Education Academy in Advanced Higher Education from the United Kingdom in 2017.

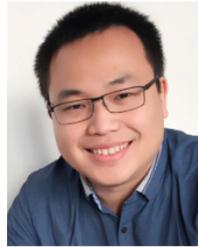

**Matthew M. Y. Kuo** (Member, IEEE) received the B.E. (Hons.) and Ph.D. degrees in electrical and computer systems engineering from the University of Auckland, Auckland, New Zealand, in 2008 and 2015, respectively.

He is currently a Lecturer with the School of Engineering, Computer and Mathematical Sciences, Auckland University of Technology. His current research interests include cyber–physical embedded systems, Internet-of-Things, robotics, precision-timed systems, industrial automation, and safety-critical systems.